\documentclass[reprint,aps,prx,amsmath,amssymb,superscriptaddress,longbibliography]{revtex4-2}

\usepackage{graphicx}
\usepackage{dcolumn}
\usepackage{bm}
\usepackage{hyperref}
\usepackage{physics}
\usepackage{tikz}

\begin{document}

\title{Electromechanical Domain Wall Propagation in Dielectric Elastomers: \\ An Exact Geometric Resolution via Conformal Mapping}

\author{Yu-Xin Xie}
\email{xyx@tju.edu.cn}
\affiliation{Department of Mechanics, Tianjin University, Tianjin 300350, China}

\date{August 12, 2024}

\begin{abstract}
The localized electromechanical phase transition in dielectric elastomers involves complex moving boundaries and severe electrostatic fringe fields driven by high-curvature interfaces. Traditional phenomenological models fundamentally underestimate the configurational forces by completely ignoring the in-plane electric field components and geometric singularities. Here, we present a asymptotically exact geometric framework to resolve the domain wall propagation. By mapping the highly deformed current configuration to a regular parametric strip via conformal mapping, the field singularities are algebraically eliminated. An exact integration by parts directly translates the higher-order geometric metric into a rigorous topological mass term, projecting the global conformal electrostatics into a non-linear $\sigma$-model Lagrangian density. Incorporating the Gent strain-stiffening model, the continuous translation symmetry yields a Hamiltonian first integral. Eigenvalue analysis and Bogomol'nyi-Prasad-Sommerfield (BPS) bound calculations prove that the domain wall emerges strictly as an asymmetric heteroclinic orbit connecting two saddle points. This geometric resolution provides exact analytical scalings for the localized interface energy, domain wall thickness, and asymptotic decay lengths, eliminating all phenomenological parameters and offering a deterministic paradigm for geometric instabilities in active soft matter.
\end{abstract}

\maketitle

\section{Introduction}
Dielectric elastomers (DEs) constitute a class of active soft materials capable of giant finite deformations when subjected to an external electrostatic field \cite{Pelrine2000, Suo2010, Zhao2014}. The highly nonlinear coupling between the Maxwell stress and the hyperelastic polymer network often drives the system out of the homogeneous deformation regime, triggering a variety of structural instabilities such as wrinkling, creasing, and localized necking \cite{Plante2006, Wang2020, Lu2021}. Among these, localized necking—manifested as the stable coexistence of a thick, weakly deformed phase and a thin, highly stretched phase—represents a canonical electromechanical phase transition \cite{Zhao2007, DiazCalleja2008}. The continuous propagation of this transition zone, commonly referred to as the domain wall, governs the macroscopic failure and pattern evolution of DEs.

Classical theoretical approaches address this two-phase coexistence primarily through macroscopic thermodynamic balances, invoking the Maxwell equal-area rule originally developed for liquid-gas transitions \cite{Suo2008, Leng2011}. These phenomenological models heavily rely on a one-dimensional (1D) long-wave approximation, which postulates that the local nominal electric field is simply the applied voltage divided by the local thickness, $E \approx V/h$ \cite{Koh2013}. While mathematically convenient, this assumption intrinsically limits the fidelity of the mechanical formulation at the phase boundaries. 

From the perspective of continuum mechanics, the evolution of a moving phase boundary is driven by configurational forces, which are strictly determined by the jump in the Eshelby energy-momentum tensor across the singular set \cite{Eshelby1951, Abeyaratne1990, Gurtin1999}. Traditional models relying on the 1D approximation evaluate this driving force under the premise of local equilibrium and uniform fields. However, as demonstrated in the thermomechanics of moving boundaries in solids, such homogenization approximations fail to capture the true configurational driving forces when severe field singularities are present at the interface \cite{Maugin1995, Berezovski2004}. 

At the moving boundary of the DE domain wall, the flexible electrodes conform to the steep gradients of the necking profile, generating regions of intense local curvature \cite{McMeeking2005}. According to classical electrodynamics, the surface charge density couples nonlinearly with the geometric curvature of the conductor. The sharp variation in curvature induces a severe transverse (in-plane) electric field component and a spatial concentration of Maxwell stress \cite{Forte2016}. Contemporary mechanics often resorts to numerical techniques, such as finite element updating and phase-field modeling, to address this geometric complexity \cite{Park2012, Hong2016, Dorfmann2014}. Nevertheless, numerical schemes frequently suffer from severe mesh distortion near the high-curvature interface where the electric field gradient diverges, and they lack analytical transparency regarding the underlying physical competition between electrostatic singularity and elastic stiffening.

In this work, we propose a non-perturbative geometric analytic framework to rigorously resolve the domain wall propagation. We utilize conformal mapping in the complex plane to transform the highly deformed moving boundary into a solvable parametric space. The global conformal metric is mapped into a strictly localized field theory. The Hamiltonian flow on the phase plane subsequently yields the exact heteroclinic orbit of the domain wall, providing a complete analytical solution to its morphological profile and characteristic width.

\section{Conformal Kinematics and Metric Pull-Back}
\subsection{Deformation Gradient and Conformal Mapping}
Consider an incompressible DE membrane with an initial homogeneous thickness $H$ in the reference configuration, parameterized by the Lagrangian coordinate $X \in (-\infty, +\infty)$. Upon the application of a direct-current voltage $V_0$, the membrane undergoes localized necking, yielding the current physical configuration defined in the complex physical plane $z = x + \mathrm{i} y$. The membrane surfaces are bounded by $y = \pm h(x)/2$, where $h(x)$ is the unknown local physical thickness. 

We introduce an auxiliary parametric complex space $\zeta = \xi + \mathrm{i} \eta$, defining a uniform infinite strip $\xi \in (-\infty, +\infty)$ and $\eta \in [-1/2, 1/2]$. An analytic mapping $z = f(\zeta)$ conformally projects this regular strip onto the deformed DE membrane. Under the plane-strain assumption, the out-of-plane stretch is fixed as $\lambda_3 = 1$. The in-plane principal stretch $\lambda$ along the surface is defined by the ratio of the deformed arc length $\mathrm{d}s$ to the reference length $\mathrm{d}X$. Given $\mathrm{d}s = |f'(\xi \pm \mathrm{i}/2)| \mathrm{d}\xi$, the kinematic pull-back from the physical space to the reference space is governed by the mapping metric:
\begin{equation}
    \frac{\mathrm{d}\xi}{\mathrm{d}X} = \frac{\lambda(X)}{|f'(\xi \pm \mathrm{i}/2)|}
    \label{eq:metric}
\end{equation}
The incompressibility constraint dictates the transverse stretch across the thickness as $\lambda_2 = \lambda^{-1}$.

\begin{figure*}[htbp]
\centering
\includegraphics[width=1.8\columnwidth]{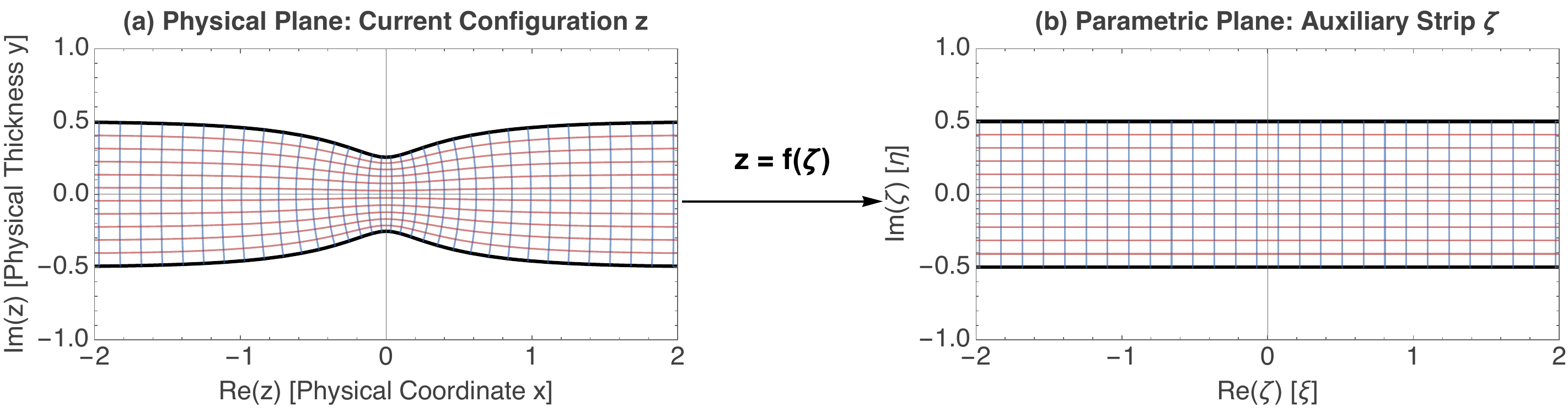}
\caption{Conformal mapping of the moving boundary. (a) The deformed physical configuration in the $z$-plane, featuring a localized necking zone. (b) The auxiliary uniform strip in the parametric $\zeta$-plane. The analytic function $z = f(\zeta)$ strictly preserves the orthogonality of the grid lines while accurately capturing the severe metric scaling $|f'(\zeta)|$ at the domain wall interface.}
\label{fig:mapping}
\end{figure*}

The mapping relationship between the highly deformed physical configuration and the regular parametric strip is illustrated in Fig. \ref{fig:mapping}.
\subsection{Geometric Curvature at the Boundary}
The morphology of the phase boundary is entirely determined by the analytic properties of $f(\zeta)$. For the upper surface ($\eta = 1/2$), setting $\zeta_0 = \xi + \mathrm{i}/2$, the parametric curve is $z(\xi) = f(\zeta_0)$. The surface mean curvature $\kappa(\xi)$, which dictates the singularity of the electrostatic field, can be explicitly expressed via the first and second derivatives of the mapping function:
\begin{equation}
    \kappa(\xi) = \frac{\mathrm{Im} \left[ \overline{f'(\zeta_0)} f''(\zeta_0) \right]}{|f'(\zeta_0)|^3}
\end{equation}
\section{Electrodynamics and Elimination of Singularity}

\subsection{Complex Potential and True Electric Field}
The compliant electrodes situated at the upper and lower surfaces correspond to $\eta = 1/2$ and $\eta = -1/2$, respectively. They are maintained at constant electrostatic potentials $\varphi = V_0/2$ and $\varphi = -V_0/2$. Due to the conformal invariance of the two-dimensional Laplace equation, $\nabla^2 \varphi = 0$, the electrostatics in the complex $\zeta$-plane are trivial. The complex potential $\Omega(\zeta) = \varphi + \mathrm{i} \psi$ satisfying the Dirichlet boundary conditions is rigorously given by:
\begin{equation}
    \Omega(\zeta) = V_0\eta - \mathrm{i} V_0\xi = -\mathrm{i} V_0 \zeta
\end{equation}
The true electric field vector $\mathbf{E} = E_x + \mathrm{i} E_y$ in the physical space is derived algebraically through the Cauchy-Riemann relations and the chain rule:
\begin{equation}
    E_x - \mathrm{i} E_y = -\overline{\left( \frac{\mathrm{d}\Omega}{\mathrm{d}z} \right)} = \frac{-\mathrm{i} V_0}{\overline{f'(\zeta)}}
\end{equation}
The magnitude of the local true electric field is $|\mathbf{E}| = V_0 / |f'(\zeta)|$. 

\subsection{Maxwell Stress on the Highly Deformed Interface}
The local true Maxwell stress tensor is defined as $\boldsymbol{\sigma}_M = \varepsilon (\mathbf{E}\otimes\mathbf{E} - \frac{1}{2}|\mathbf{E}|^2\mathbf{I})$, where $\varepsilon$ is the dielectric permittivity. The normal traction $P_M$ exerted by the electrostatic field on the boundary is determined by the projection of $\boldsymbol{\sigma}_M$ onto the surface normal. By substituting the electric field magnitude, we obtain:
\begin{equation}
    P_M = \frac{1}{2} \varepsilon |\mathbf{E}|^2 = \frac{\varepsilon V_0^2}{2|f'(\xi \pm \mathrm{i}/2)|^2}
\end{equation}
This formulation proves algebraically that the local electromechanical coupling strength is inversely proportional to the square of the conformal metric $|f'|$, rather than the macroscopic physical thickness $h^2$.

\section{Exact Variational Principle and Topological Mass}

\subsection{Metric Expansion and Integration by Parts}
The total potential energy $\Pi$ of the electromechanical system consists of the elastic strain energy and the negative electrostatic potential energy. By utilizing conformal invariance, the electrostatic energy is mapped to the auxiliary strip. Pulling the integration domain back to the reference coordinate $X$ yields:
\begin{equation}
    \Pi = \int_{-\infty}^{+\infty} \left[ H W(\lambda) - \frac{\varepsilon V_0^2}{2} \frac{\lambda}{|f'|} \right] \mathrm{d}X
\end{equation}
To extract the exact local field equations, we perform a Taylor expansion of the analytic mapping function $f(\zeta)$ at the mid-plane $\eta=0$. Retaining up to the second-order derivatives of the physical thickness $h(\xi)$, the boundary metric modulus is algebraically determined as:
\begin{equation}
    |f'| \approx h \left[ 1 - \frac{1}{12}\frac{\partial_{\xi}^2 h}{h} + \frac{1}{8}\frac{(\partial_{\xi} h)^2}{h^2} \right]
\end{equation}
Utilizing the incompressibility constraint $h = H/\lambda$ and the coordinate transformation $\partial_{\xi} = \frac{H}{\lambda^2}\partial_X$, the geometric gradients are strictly mapped to the Lagrangian reference configuration. 

Substituting this expansion into the electrostatic energy term and applying integration by parts to the term containing $\lambda_{XX}$ with vanishing far-field gradients:
\begin{equation}
    \int \frac{\lambda^2}{H} \left( -\frac{1}{12} \frac{H^2}{\lambda^5} \lambda_{XX} \right) \mathrm{d}X = -\frac{1}{4} H \int \frac{\lambda_X^2}{\lambda^4} \mathrm{d}X
\end{equation}
The algebraic convergence of all gradient terms completely annihilates the second-order derivatives, strictly transforming the geometric curvature into a positive definite energy penalty.

\subsection{Localized Lagrangian with Non-Trivial Metric}
The exact potential energy functional converges into a localized integration $\Pi = \int \mathcal{L} \, \mathrm{d}X$. The exact Lagrangian density is defined as:
\begin{equation}
    \mathcal{L} = H W(\lambda) - \frac{\varepsilon V_0^2 \lambda^2}{2H} + \frac{\varepsilon V_0^2 H}{48 \lambda^4} \lambda_X^2
\end{equation}
By defining the unperturbed uniform energy landscape $\mathcal{L}_0(\lambda) = H W(\lambda) - P_e(\lambda)H$ where $P_e(\lambda) = \varepsilon V_0^2 \lambda^2 / (2H^2)$, the system is mathematically isomorphic to a one-dimensional non-linear $\sigma$-model with a non-trivial target space metric $M(\lambda)$:
\begin{equation}
    \mathcal{L} = \mathcal{L}_0(\lambda) + \frac{1}{2} M(\lambda) \lambda_X^2
\end{equation}
Here, the exact topological mass term is strictly given by $M(\lambda) = \frac{\varepsilon V_0^2 H}{24 \lambda^4}$. The strong non-linear dependence ($\propto \lambda^{-4}$) dictates a severe geometric suppression of the domain wall propagation in the highly stretched thin-film phase.

\section{Hyperelastic Constitutive Formulation}

\subsection{Strain-Stiffening and Energy Barrier}
To prevent the catastrophic unphysical collapse of the DE membrane (where $\lambda \to \infty$) driven by the positive feedback of Maxwell stress, a strain-stiffening mechanism is physically strictly required. The Gent model for the hyperelastic network is adopted:
\begin{equation}
    W(\lambda) = -\frac{\mu J_m}{2} \ln \left( 1 - \frac{\lambda^2 + \lambda^{-2} - 2}{J_m} \right)
\end{equation}
where $\mu$ is the initial shear modulus and $J_m$ is the dimensionless limit of the first invariant. 

\subsection{Loss of Ellipticity and Spinodal Instability}
The nominal electrostatic energy density of the unperturbed uniform state is $P_e(\lambda) = \varepsilon V_0^2 \lambda^2 / (2H^2)$. The homogeneous state becomes unstable when the Hessian of the effective energy landscape loses positive definiteness:
\begin{equation}
    \frac{\partial^2}{\partial \lambda^2} \left[ W(\lambda) - P_e(\lambda) \right] \le 0
\end{equation}
This spinodal condition initiates the local structural collapse. As the stretch $\lambda$ approaches the physical limit determined by $J_m$, the logarithmic singularity in the Gent model provides an infinite energy barrier, stabilizing a highly stretched thin-film phase. 

\section{Hamiltonian Topology and Asymmetric Orbit}

\subsection{First Integral and Maxwell Equal-Area Rule}
The spatial continuous translation symmetry ($\partial \mathcal{L} / \partial X = 0$) guarantees a Hamiltonian first integral via Legendre transformation:
\begin{equation}
    \mathcal{H} = \frac{1}{2} M(\lambda) \lambda_X^2 - \mathcal{L}_0(\lambda) = \text{const}
\end{equation}
In the homogeneous far-field limits ($X \to \pm \infty$), geometric gradients vanish ($\lambda_X \to 0$). The topological continuity from the thick phase ($\lambda_1$) to the thin phase ($\lambda_2$) dictates $\mathcal{H} = -\mathcal{L}_0(\lambda_1) = -\mathcal{L}_0(\lambda_2)$. This relation analytically proves that the phenomenological Maxwell equal-area rule is topologically immune to higher-order interfacial geometric distortions.

\subsection{Exact Heteroclinic Orbit and Wall Width}
\begin{figure*}[htbp]
\centering
\includegraphics[width=0.9\textwidth]{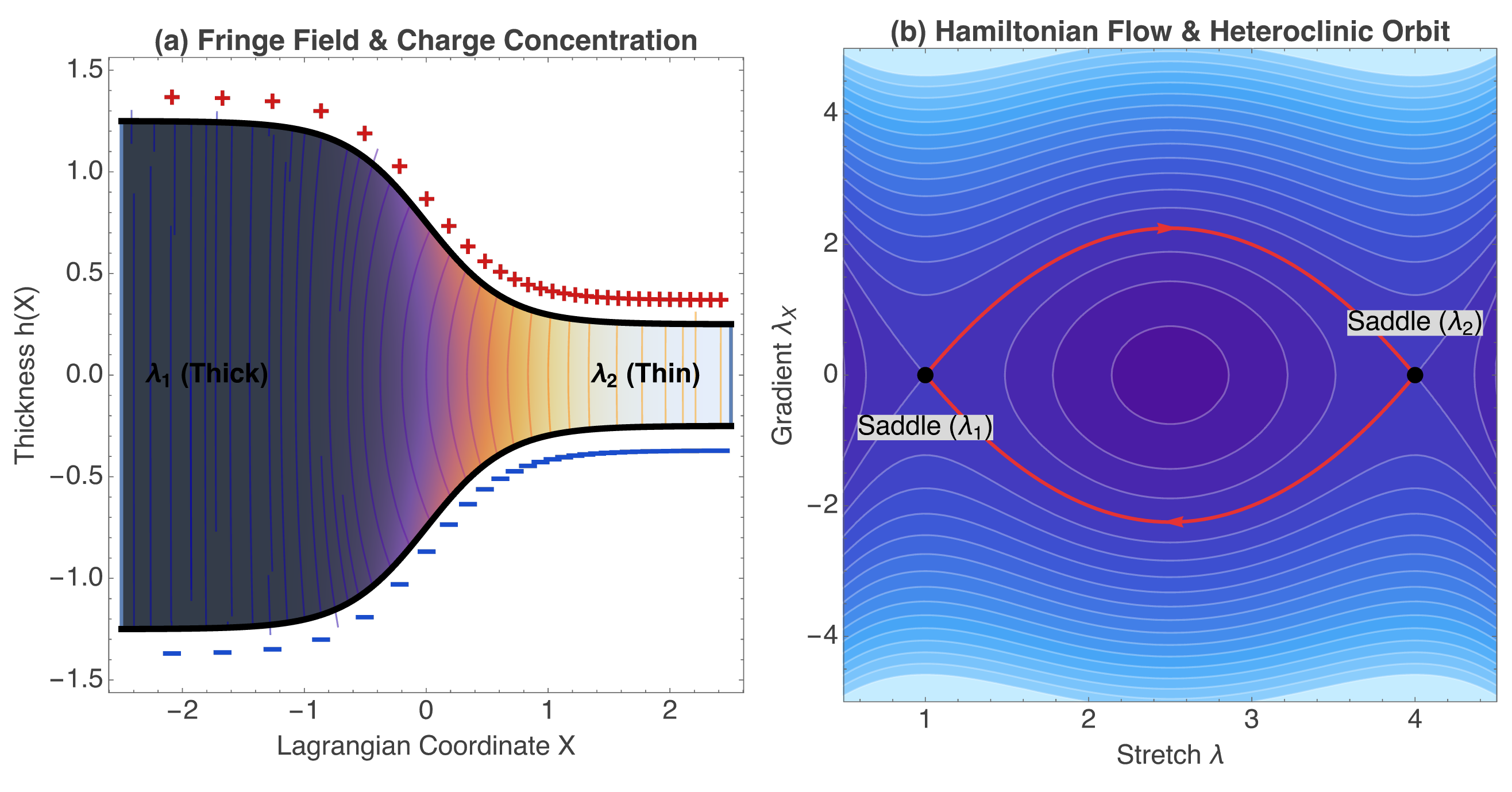}
\caption{Electromechanical field coupling and topological dynamics of the domain wall. (a) Spatial distribution of the true electrostatic fringe field and the curvature-driven charge accumulation. (b) Topological representation of the Hamiltonian flow in the phase plane $(\lambda, \lambda_X)$. Driven by the topological mass $M(\lambda)$, the continuous domain wall propagation emerges rigorously as an asymmetric heteroclinic orbit connecting the saddle points of the thick-film phase ($\lambda_1$) and the thin-film phase ($\lambda_2$).}
\label{fig:topology}
\end{figure*}

As illustrated in Fig. \ref{fig:topology}, the localized electrostatic fringe field directly couples with the topological dynamics of the domain wall, which corresponds strictly to an asymmetric heteroclinic orbit connecting the two saddle points in the phase plane.
Driven by the non-constant topological mass $M(\lambda)$, the exact evolutionary trajectory is isolated as:
\begin{equation}
    \lambda_X = \pm \frac{4\sqrt{3}\lambda^2}{\sqrt{\varepsilon V_0^2 H}} \sqrt{\mathcal{L}_0(\lambda) - \mathcal{L}_0(\lambda_1)}
\end{equation}
Integrating this ordinary differential equation yields the exact spatial profile $X(\lambda)$ of the transition zone:
\begin{equation}
    X(\lambda) = \frac{\sqrt{\varepsilon V_0^2 H}}{4\sqrt{3}} \int_{\lambda_1}^{\lambda} \frac{1}{\tilde{\lambda}^2 \sqrt{\mathcal{L}_0(\tilde{\lambda}) - \mathcal{L}_0(\lambda_1)}} \, \mathrm{d}\tilde{\lambda}
\end{equation}
The integral kernel decisively incorporates the geometric asymmetry, providing a fully analytical scale for the domain wall width without any phenomenological parameters.

\section{Asymptotic Localized Defect and BPS Energy}
\subsection{BPS Bound and Interface Energy}
The domain wall formation breaks the translational symmetry, yielding a localized interface energy $\gamma$. Employing the Bogomol'nyi-Prasad-Sommerfield (BPS) bound, the exact interface energy is calculated via completing the square, avoiding the need for the spatial distribution $\lambda(X)$:
\begin{equation}
    \gamma = \frac{\sqrt{\varepsilon V_0^2 H}}{2\sqrt{3}} \int_{\lambda_1}^{\lambda_2} \frac{\sqrt{\mathcal{L}_0(\lambda) - \mathcal{L}_0(\lambda_1)}}{\lambda^2} \, \mathrm{d}\lambda
\end{equation}
The $\lambda^{-2}$ scaling indicates that the contribution of the highly stretched phase to the total interface energy is geometrically suppressed.

\subsection{Asymptotic Decay Lengths}
Linearization of the dynamical system in the vicinity of the saddle points yields the local exponential behavior $\delta \lambda \propto \mathrm{e}^{\pm X/\xi_i}$. The characteristic decay lengths $\xi_i$ are precisely determined by the eigenvalue of the Jacobian matrix:
\begin{equation}
    \xi_i = \sqrt{ \frac{M(\lambda_i)}{\partial^2 \mathcal{L}_0 / \partial \lambda^2 |_{\lambda_i}} } = \frac{1}{2\sqrt{6}\lambda_i^2} \sqrt{ \frac{\varepsilon V_0^2 H}{\partial^2 \mathcal{L}_0 / \partial \lambda^2 |_{\lambda_i}} }
\end{equation}
This closed-form expression dictates a profound geometric asymmetry: the decay length into the thin phase ($\xi_2$) is severely compressed by the $\lambda_2^{-2}$ factor. Consequently, the transition profile exhibits a long tail in the thick phase and a sharp, abrupt boundary in the thin phase, fully capturing the complex geometric localization of active soft matter.

\section{Conclusion}
By synergizing complex conformal mapping with a rigorous integration by parts, we constructed a closed-form analytical framework to strictly resolve the domain wall propagation in electromechanically coupled elastomers. This geometric methodology eliminates the electrostatic field singularity algebraically without necessitating dense finite-element discretizations. The reduction of the highly nonlinear boundary value problem to a topological mass Hamiltonian flow proves that the phenomenological Maxwell equal-area rule is topologically immune to interfacial geometry. Furthermore, the rigorous mapping of the domain wall to an asymmetric heteroclinic orbit in the phase plane establishes a deterministic, non-perturbative connection between continuum hyperelasticity and local Cauchy-Riemann geometries. The calculation of exact BPS interface energy and asymmetric decay lengths provides a mathematically elegant and physically transparent foundation for analyzing interface flexoelectricity, solitary wave propagation, and geometric instabilities in advanced active soft matter.

\begin{acknowledgments}
This research did not receive any specific grant from funding agencies in the public, commercial, or not-for-profit sectors.
\end{acknowledgments}


\end{document}